# Spontaneous generation and active manipulation of real-space optical vortex

Dongha Kim[1,#], Arthur Baucour[2], Yun-Seok Choi[3], Jonghwa Shin[2], Min-Kyo Seo[1,*]

[1]Department of Physics, KAIST, Daehak-ro, 291, 34141, Republic of Korea

[2]Department of Materials Science and Engineering, KAIST, Daehak-ro, 291, 34141, Republic of Korea

[3]Department of Chemistry, KAIST, Daehak-ro, 291, 34141, Republic of Korea

[#]dongha_kim@kaist.ac.kr

[*]minkyo_seo@kaist.ac.kr

**Abstract**

**Optical vortices host the orbital nature of photons[1], which offers an extra degree of freedom in photonic applications[2,3,4,5,6,7,8]. Unlike vortices in other physical entities, optical vortices require structural singularities[9,10,11,12], which restrict their abilities in terms of dynamic and interactive characteristics[13,14,15,16,17]. In this study, we present the spontaneous generation and external magnetic field-induced manipulation of an optical vortex and antivortex. A gradient-thickness optical cavity (GTOC) consisting of an Al/SiO$_2$/Ni/SiO$_2$ multilayer structure realised the distinct transition between the trivial and non-trivial topological phases, depending on the magneto-optic effects of the Ni layer. In the non-trivial topological phase, the mathematical singularities generating the optical vortex and antivortex pair in the reflected light existed in the generalised parameter space of the thicknesses of the top and bottom SiO$_2$ layers, which is bijective to the real space of**

**the GTOC. Coupled with the magnetisation, the optical vortex and antivortex in the GTOC experienced an effective spin-orbit interaction and showed topology-dependent dynamics under external magnetic fields. We expect that field-induced engineering of optical vortices will pave the way for the study of topological photonic interactions and their applications.**

**Introduction**

Vortices, supporting non-trivial topological textures with singularities, have been demonstrated in various physical entities[18,19]. In condensed matter, vortices emerge from exotic physical phenomena, such as dipolar interaction in spintronics[20,21], polariton condensation in excitonics[22,23], and magnetic flux quantisation in type-II superconductors[24,25]. Observation of their binding and unbinding phenomena, such as the Berezinskii-Kosterlitz-Thouless transition[26,27], introduced a new era in physics, which explored newly emerging phases of matter on account of the interactions between topological textures. The optical vortex (OV) is a topological texture in electromagnetic fields[1] having a zero-intensity singular point and a spiral phase front that has attracted immense attention in optical analogous to the vortices of other physical entities. OVs have been demonstrated in a variety of forms, from free-propagating waves[28] to evanescent fields[29]. Supporting orbital angular momentum, OVs have been widely employed in fundamental studies and applications, such as spin-orbit interaction-mediated manipulation[2,3,4], forbidden electronic transition[5], classical/quantum information processing[6,7], and super-resolution microscopy[8].

However, conventional OVs have been implemented by employing fixed structural singularities, such as spiral phase plates[9], phased array antennas[10], and spatially engineered meta-surfaces[11,12], rather than critical phenomena. The employment of fixed structural singularities prevents OVs from possessing quasi-particle-like behaviour, such as dynamic and interactive characteristics[13,14,15,16,17]. Recent studies reported on the structural singularity-free generation of OVs, employing the bound states in the continuum of photonic crystals[30,31] and the photorefractive response in a nonlinear bulk medium[32], however, they were either implemented in the momentum space or were inappropriate for on-demand, external control. On the other hand, the topological vortices formed spontaneously in condensed matter systems have taken advantage of spatially flexible and externally controllable behaviours. Thus, providing a platform for structural singularity-free generation and dynamic manipulation of real-space OVs will enables a new way to investigate topological interactions in photonics, including topological Hall effect[33,34], chiral spin-orbit coupling[35,36], and topological phase transition[37,38], potentially leading to developments as rich as the condensed-matter vortices. In this study, we demonstrated the spontaneous generation of real-space optical vortex and antivortex and their active control using an external magnetic field, at the heart of which is a gradient-thickness optical cavity (GTOC).

The GTOC consisted of a magneto-optic Ni layer, between two $SiO_2$ layers, on top of an Al mirror, as shown in Figure 1a. The thicknesses of the top and bottom $SiO_2$ layers, $h_1$ and $h_2$ respectively, gradually changed in almost orthogonal directions over the GTOC. The spatial area of the GTOC was mapped bijectively onto a part of the generalised parameter space of ($h_1$, $h_2$). The optical thickness of the Ni layer, which can be controlled by an external magnetic field, characterised the topological texture of the reflected light from the GTOC. For most of the thickness range of the Ni layer, the GTOC exhibited a weak, non-zero minimum reflectance as a function of the thicknesses of the top and bottom $SiO_2$ layers (Figure 1b). Enclosing the weak minimum, the texture of the complex reflection has zero topological charge ($w = 0$) and does not contain any notable features. We named this typical case, the trivial topological phase. On the other hand, for a specific range of Ni layer thicknesses, the GTOC exhibited a unique topological phase in reflection, that we called, the non-trivial topological phase. As shown in Figure 1c, the non-trivial topological phase possesses two singular points of zero reflection in the generalised parameter space, and their phase distributions support the optical vortex ($w = +1$) and antivortex ($w = -1$), respectively. This new type of singularity originated from the mathematical topology around the zero-reflection solution in the generalised parameter space, and completely replaced the conventional structural singularity required to generate optical vortex and antivortex. Moreover, the structural freedom of the GTOC in the topological singularity generation made it possible to actively create, annihilate, and transport the optical vortex and antivortex by an external stimulus, that is, the applied magnetic field, as if they were quasi-particles.

**Non-trivial topological phase of GTOC**

The topological texture in the GTOC is periodically repeated depending on the thickness of the top and bottom $SiO_2$ layers, as shown in the plots of the reflection amplitude and phase in the generalised parameter space (Figure 1d). The periodicity ($p$) is given as $\lambda/2n_{SiO2}$, where $\lambda$ and $n_{SiO2}$ are the wavelength of light (632.8 nm) and the refractive index of $SiO_2$ (1.472), respectively. When the thickness of the Ni layer ($h_{Ni}$) was 5 nm, a weak reflection minimum was observed per unit cell ($p \times p$) in the generalised parameter space, without a phase singularity. The 5-nm-thick Ni layer was optically thin so that the reflection phase depended on the thickness of both the top and bottom $SiO_2$ layers. The GTOC with $h_{Ni} = 15$ nm also exhibited a single weak reflection minimum in the unit cell; however, because of the optically thick Ni layer, the thickness of the top $SiO_2$ layer dominated the tendency of the reflection phase change. On the other hand, the GTOC of $h_{Ni} = 10$ nm supported two singular points of extremely suppressed reflection in the unit cell, which possessed opposite topological charges

($w = \pm1$), and created a pair of optical vortex and antivortex. The Ni layer with an appropriate thickness allowed complete destructive interference between the optical fields in the top and bottom SiO$_2$ layers, which is the necessary condition for the singularity of zero reflection. Figure 1e plots the position and topological charge of the weak and singular minima of reflection in the generalised parameter space.

The solution of the singular zero-reflection ($r = 0$) in the generalised parameter space of the GTOC exists within a certain range of the Ni layer thicknesses. The multiple beam interference theory formulates the reflection coefficient by combining two participating terms: 1) optical paths including the propagations only above the Ni layer and 2) the other paths including the transmission through the Ni layer (Supplementary Note S1). The interference of the two terms enables us to analyse and understand the creation of the non-trivial topological phase. The singular minimum of reflection exists only when the two terms have the same amplitude and are out of phase. Each requirement appears as a closed-loop (same amplitude condition) and finite curved lines (out-of-phase condition) in the generalised parameter space, resulting in a paired solution for the zero-reflection at their intersections. Intersection of the same amplitude and out-of-phase condition was established only between two critical thicknesses of the Ni layer, $h_{c1} = 6.29$ nm and $h_{c2} = 13.15$ nm.

The evolution of the reflectance minimum and topological charge depending on the Ni layer thickness confirmed the topological phase transition of the GTOC (Figure 1f). Here, the topological charge was determined by integrating the reflection phase along a closed-path, enclosing the reflectance minimum in the generalised parameter space. For $h_{Ni} < h_{c1}$ ($h_{Ni} > h_{c2}$), a single reflectance minimum was observed, accompanied by the zero topological charge, and its value gradually decreased from $7.77 \times 10^{-1}$ to $6.56 \times 10^{-5}$ (from $5.78 \times 10^{-2}$ to $5.60 \times 10^{-6}$) as the Ni layer thickness approached the critical thickness. In the region of $h_{c1} < h_{Ni} < h_{c2}$, two reflection minima with opposite, non-trivial topological charges of +1 and −1 were observed, and their values in the order of $10^{-7}$ were three orders of magnitude smaller than that of the trivial weak reflection minimum. The transition between the trivial and non-trivial topological phases exhibited a discontinuity in the first derivative of the reflectance and topological charge value at the critical thicknesses, which enabled us to realise the distinct, and abrupt creation/annihilation of the optical vortex-antivortex pair by the magneto-optic effects on the Ni layer (Figure 4). It was also guaranteed that throughout the regime of the non-trivial topological phase, an external magnetic field dynamically controlled the behaviour of the optical vortex and antivortex (Figure 3).

**Demonstration of real-space OV at non-trivial topological phase**

We experimentally demonstrated the trivial and non-trivial topological phases in the GTOCs with different Ni layer thicknesses (Figure 2). We fabricated the samples on a square piece of Si substrate, by sequential deposition of the Al mirror, SiO$_2$ first layer ($h_2$), Ni, and SiO$_2$ second layer ($h_1$) (Figure 2a). The SiO2 layers were deposited by radio-frequency (RF) sputtering, which exhibits directional deposition. Therefore, the SiO$_2$ layers exhibited a gradual change in thickness over the sample, with the direction and range of the thickness gradient depending on the orientation and position of the sample with respect to the centre of the deposition pattern. Further information about the thickness distribution and gradient in the SiO$_2$ can be found in Supplementary Note 2. The Ni layer was deposited uniformly over the sample area (See Methods). The thickness gradients of the top and bottom SiO$_2$ layers ($\nabla h_1$ and $\nabla h_2$) were engineered to be almost perpendicular, and their magnitudes were in the order of nm/mm. Under these conditions, the fabricated 2×2-cm$^2$-sized sample covered a partial area of the GTOC unit cell in the generalised parameter space (indicated region by the yellow line in

Figure 2b), that contained a weak or singular minimum of reflection. The bijective projection between the real and generalised parameter spaces was performed by considering the spatially varying magnitudes of the thickness gradients over the samples. As shown in Figure 2b, we examined four different samples (namely, Sample #1, #2, #3, and #4) supporting the trivial weak minimum and the vortex/antivortex singularity predicted in Figures 1d and 1e.

In Figure 2c, we measured the reflectance of the GTOCs depending on their position using a confocal microscope. The GTOCs of $h_{Ni}$ = 5 (Sample #1) and 15 nm (Sample #4) generated a weak reflection minimum with a reflectance of approximately $2.46 \times 10^{-2}$ and $3.00 \times 10^{-2}$, respectively. Their reflectance distributions showed the behaviour of a normal differentiable function in two dimensions. On the other hand, the GTOCs of $h_{Ni}$ = 10 nm clearly produced a singular point in the reflectance distribution: $|r|^2$ = ~$4.04 \times 10^{-5}$ (Sample #2) and ~$2.30 \times 10^{-4}$ (Sample #3). As previously mentioned, the existence of a non-differentiable reflectance extremum is necessary to support the phase singularity for an optical vortex or antivortex. The amplitude and phase distribution of the complex-valued reflection coefficient, $r(x, y)$, directly revealed the topological texture created by the GTOC (Figure 2d). We measured the complex reflection coefficient distribution using off-axis holography (see Methods and Supplementary Note S3). Samples #1 and #4 exhibited trivial topology without any phase singularity. However, Samples #2 and #3 built counter-clockwise and clockwise phase windings, corresponding to the optical vortex and antivortex, respectively, without any structural singularity. The footprint of the optical vortex and antivortex can be reduced at the micrometre-scale by increasing the thickness gradients of the GTOC.

**Real-space dynamics of OVs under effective spin-orbit interaction**

The GTOC activated topology-dependent magneto-optic dynamics of the optical vortex and antivortex, interacting with the external magnetic field (Figure 3a). The off-diagonal term of the permittivity tensor, which is related to the Voigt parameter, sensitively responded to the induced magnetisation (**M**) in the Ni layer and caused a magnetic circular birefringence. The magnetic birefringence enabled the engineering of the optical thickness of the Ni layer for the circularly-polarized light. The coupling between the out-of-plane magnetisation of the Ni layer (**M** ∥ **ẑ**) and the topological charge of the optical field (**w** = $w\hat{z}$, $w = -1, 0, +1$) governed the dynamic behaviour and movement of the topological texture. This is analogous to the spin-orbit interaction in condensed matter systems, considering the magnetisation and topological charge as the spin and orbital quantity, respectively. As the optical thickness of the Ni layer increased, the optical vortex and antivortex exhibited opposite movements, which represented the positive and negative effective spin-orbit interaction, respectively. The vortex and antivortex mainly moved along the trajectory of the singular minimum solution depending on the thickness of the Ni layer in the generalised parameter space (Figure 3b). As the magnetisation changed from the negative to positive direction, $h_{Ni}$ effectively increased from $h_{c1}$ to $h_{c2}$ (decreased from $h_{c2}$ to $h_{c1}$), for the right-handed (left-handed) circularly-polarised light. In the generalised parameter space, the optical vortex and antivortex arose at the critical thickness ($h_{c1}$) of the Ni layer, moved in opposite directions, and disappeared at the other critical thickness ($h_{c2}$), and vice versa.

Figures 3c and 3d show the behaviours of the optical vortex (Sample #2) and antivortex (Sample #3), depending on the external magnetic field for the right-handed (+σ) circularly-polarised light. While the external magnetic field changed from −0.5 T to +0.5 T, the optical vortex and antivortex showed movements in the real space as ($\Delta x$, $\Delta y$) = (0.649 mm, −0.344 mm) and (−0.385 mm, 0.010 mm), which can be projected onto the generalized parameter space as ($\Delta h_1$, $\Delta h_2$) = (0.90 nm, −0.58 nm) and (−1.03 nm, 0.18 nm), respectively, as shown in Figure 3e and 3f. It was experimentally confirmed that the effective spin-orbit interaction distinguished the dynamics of the optical vortex and antivortex. The effective change of $h_{Ni}$ by the magnetic field were estimated to be ~0.11 nm in Sample #2 and #3, by fitting the vortex/antivortex movement to the theoretical calculation in Figure 3b. On the other hand, the trivial topological textures in Samples #1 and #4 were barely affected by the external magnetic field (Supplementary Note S4). It is worth noting that the consideration of imaginary component in the magneto-optic parameter of Ni offers precise analysis on the direction of the vortex/antivortex movement (Supplementary Note S5).

**Magnetic-field induced generation of optical vortex-antivortex pair**

Lastly, we demonstrated the magnetically-induced generation and manipulation of the optical vortex-antivortex pair in the GTOC. We fabricated a GTOC of $h_{Ni}$ = 13 nm (Sample #5) which was close to the critical thickness ($h_{c2}$ = 13.15 nm), where the transition between the non-trivial and trivial topological phases occurs. Sample #5 covered almost the same area in the generalised parameter space as the Sample #4 (Figure 2b). Figures 4a and 4b show the reflection amplitude and phase distribution of Sample #5, depending on the external magnetic field ($B$) along the z-axis. Upon applying a continuously varying global phase mathematically (Supplementary Note S6), the phase distribution in Figure 4b demonstrated the fork pattern at the location of the OVs. From $B$ = −0.5 T to 0.0 T, the GTOC supported a single trivial reflection minimum and did not exhibit a fork-like pattern in the interferogram. As the external magnetic field changed from negative to positive, the optical thickness of the Ni layer decreased, and reached the critical thickness for the topological phase transition. At $B$ = 0.1 T, the reflection amplitude was extremely suppressed, and a set of two fork patterns appeared in the holographic interferogram. The two fork patterns (red and blue dashed lines in Figure 4b) with opposite winding directions confirmed the presence of the optical vortex ($w = +1$), and antivortex ($w = -1$), respectively. As the magnetic field further increased from 0.1 to 0.5 T, the fork patterns moved in opposite directions, and separate away.

Figure 4c schematically summarises the experimental measurements of the magnetically-induced generation and manipulation of the optical vortex-antivortex pair in the GTOC. The measured trajectory along which the optical vortex and antivortex moved (dashed line in Figure 4c) matched the theoretical prediction at $h_{Ni} = h_{c2}$ (Figure 3b). According to the effective spin-orbit interaction, the optical vortex and antivortex moved ~1.20 mm and ~1.17 mm, respectively from the generation centre, in opposite directions which correspond to the displacement ($\Delta h_1$, $\Delta h_2$) of (−2.23 nm, −0.72 nm) and (2.16 nm, 0.82 nm), respectively, in the generalised parameter space. Figure 4d shows the topological transition of the GTOC depending on the external magnetic field, from −0.5 T to 0.5 T. Due to the paramagnetic behaviour of the Ni layer, the creation and annihilation of the optical vortex-antivortex pair in the current GTOC was reversible. It is worth noting that the employment of the ferromagnetic, out-of-plane magnetised layers[39,40] (Pt/Co, Pt/Co/Pt) will provide the GTOC with the ability for hysteretic effective spin-orbit interaction and topological information storage.

**Discussion**

In summary, we experimentally demonstrated the spontaneous generation and magnetic-field-induced manipulation of an optical vortex-antivortex pair in a GTOC. The spontaneous generation/annihilation of the optical vortex and antivortex originated from the mathematical condition supporting the singular zero-reflection in the generalised parameter space of the GTOC. Depending on the thickness of the Ni layer, the optical coupling of the top and bottom dielectric layers determined the presence or absence of the singularity. The magneto-optic effect on the Ni layer thus governed the abrupt, discontinuous transition between the trivial and non-trivial topological phases of the reflected light at the critical thickness. We also found the effective spin-orbit interaction between magnetisation and the optical vortex/antivortex, and realised its active control. It is worth noting that the dynamic and interactive characteristics in our system can be realised by employing a variety of other sources, including ferromagnetism[39,40,41], electro-/thermo-optic effects[42,43], and optical nonlinearities[44,45]. In this light, spontaneously generated OVs can provide a distinct platform for investigating optical analogues of the chiral spin interaction[35,36] and superfluidity[46,47] by the condensed matter counterparts. We expect the further development of our methodology will be utilised to generate higher-order topological textures[48,49], spatio-temporal electromagnetic singularities[50,51], and optical polarisation/vortex knots[52,53,54].

**Methods**

**Sample Fabrication**

A 100-nm-thick Al mirror was prepared on a 2 cm × 2 cm Si substrate by thermal evaporation. The height varying $SiO_2$ layers were achieved by radio-frequency sputtering, taking advantage of the directionality of the deposition process. To maximize the $SiO_2$ thickness range over the sample, the substrate was placed 3 cm away from the centre of the sample holder. The sample was rotated by 90° between the depositions of the top and bottom $SiO_2$ layers, so that the thickness gradients of the two layers were almost perpendicular. A magnetic Ni layer with uniform thickness was deposited using electron-beam evaporation.

**Measurement Set-Up**

The confocal microscope scanned an area of 9 mm × 9 mm, by employing a motorised XYZ stage. We employed reflective optical elements, including a right-angle metal prism mirror for beam splitting, and a concave metal mirror for beam focusing, to minimise undesired scattering/back-reflection, and improve the measurement precision. The optical readout system employing an femtowatt photoreceiver (Newport 2151), and a high-precision data acquisition tool (NI-DAQ PCIe-6321), provided a sufficiently high dynamic range to measure the reflection singularities. The off-axis holography set-up was based on an interferometric imaging system. The propagation direction of the reference beam with a 1°-tilt interfered with the signal beam reflected from the GTOC. The numerical post-process filtering in the Fourier space retrieved the complex reflection coefficient distribution from the acquired interference pattern (Supplementary Note S3). The magneto-optic effects on the Ni layer were induced using a permanent neodymium magnet. A motorised translation stage moved the magnet, and controlled the applied magnetic field to the sample.

## Supplementary Information

**Supplementary Movie 1** Plot of the same amplitude (the yellow curve) and out-of-phase (the green curve) conditions inside the GTOC unit cell ($p \times p$) in the generalised parameter space depending on the thickness of the Ni layer. The red and blue dots indicate the intersections for the optical vortex (w = +1) and antivortex (w = −1), respectively.

## Data Availability

The data that support the findings of this study are available from the corresponding author upon reasonable request.

## Acknowledgement

We thank Dr. Shinho Lee, Dr. Donghyeong Kim, Mr. Sanghyeok Park, and Ms. Hanbit Oh for their valuable discussions.

M.-K.S. acknowledges the support of the KAIST Cross-Generation Collaborative Lab project and the National Research Foundation of Korea (NRF) (2020R1A2C2014685 and 2020R1A4A2002828). J. S. acknowledges the support of the NRF (2021R1A2C200868711). D.K. acknowledges the support of the NRF (2015H1A2A1033753). Y.-S.C. acknowledges the support of the NRF (2020R1I1A1A01069219).

## Author Contribution

D. K. conceived the idea. A. B. and D. K. designed and fabricated the gradient-thickness optical cavity samples. D. K. and Y.-S. C. built the magneto-optic off-axis holography setup. D. K.

performed the magneto-optic measurements and the theoretical calculations. D. K. and M.-K.S. analysed the data and wrote the manuscript with input from all other authors.

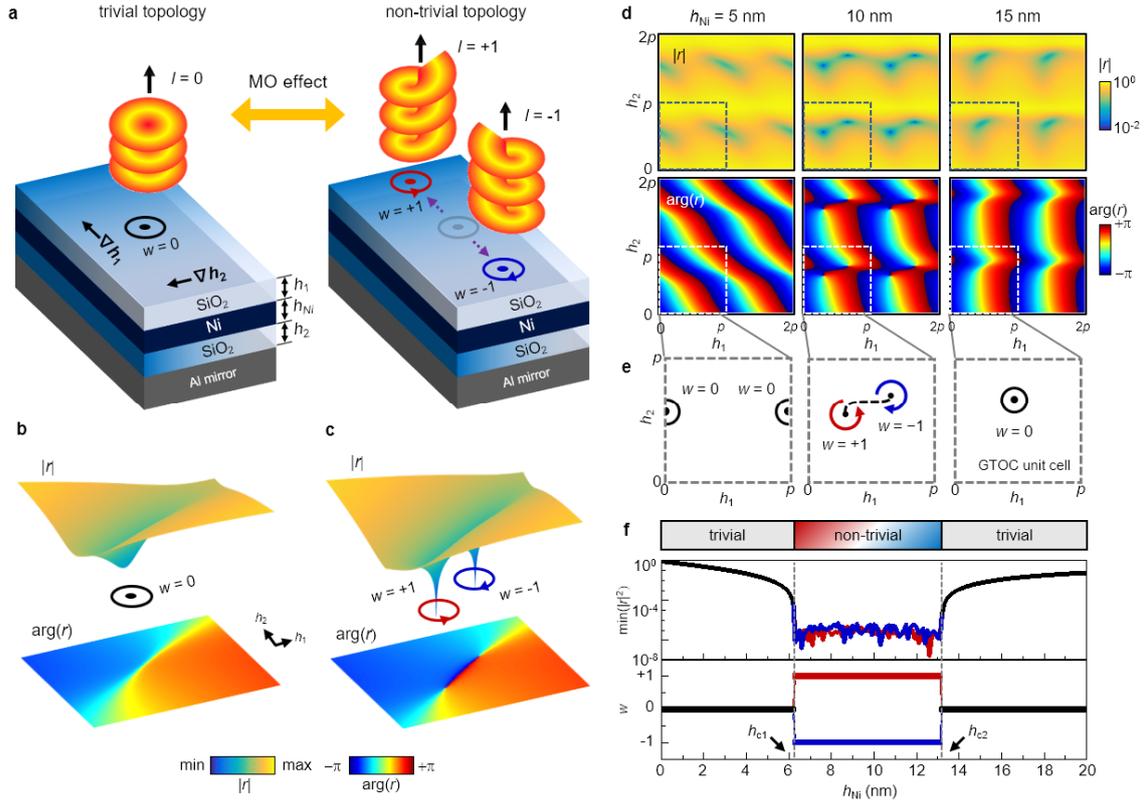

**Figure 1. Magnetically-controllable, spontaneous generation of optical vortex-antivortex pair in gradient-thickness optical cavity (GTOC).** (a) Schematic illustration of trivial and non-trivial topological phases in the GTOC. The thicknesses of the top and bottom SiO$_2$ layers ($h_1$ and $h_2$) change in orthogonal directions. Changing the optical thickness of the Ni layer varies the magneto-optic effect which governs the transition between the trivial and non-trivial topological phases. (b,c) Topological characteristics of the (b) single weak minimum, and (c) singular vortex-antivortex pair. (d) Calculated complex reflection coefficient distribution of the GTOCs with different Ni layer thicknesses ($h_{Ni}$ = 5, 10, and 15 nm) in the generalized parameter space ($h_1$, $h_2$). The topological texture repeats with a period of $p = \lambda/2n_{SiO2}$, and the dashed box indicates the GTOC unit cell. (e) Description of the topology of the reflection minimum in the GTOCs of (d). The black dashed line represents the branch cut between the vortex (red circular arrow), and antivortex (red circular arrow) singular points. (f) Calculated reflectance (min($|r|^2$)), and topological charge ($w$) of the reflection minima depending on the Ni layer thickness ($h_{Ni}$).

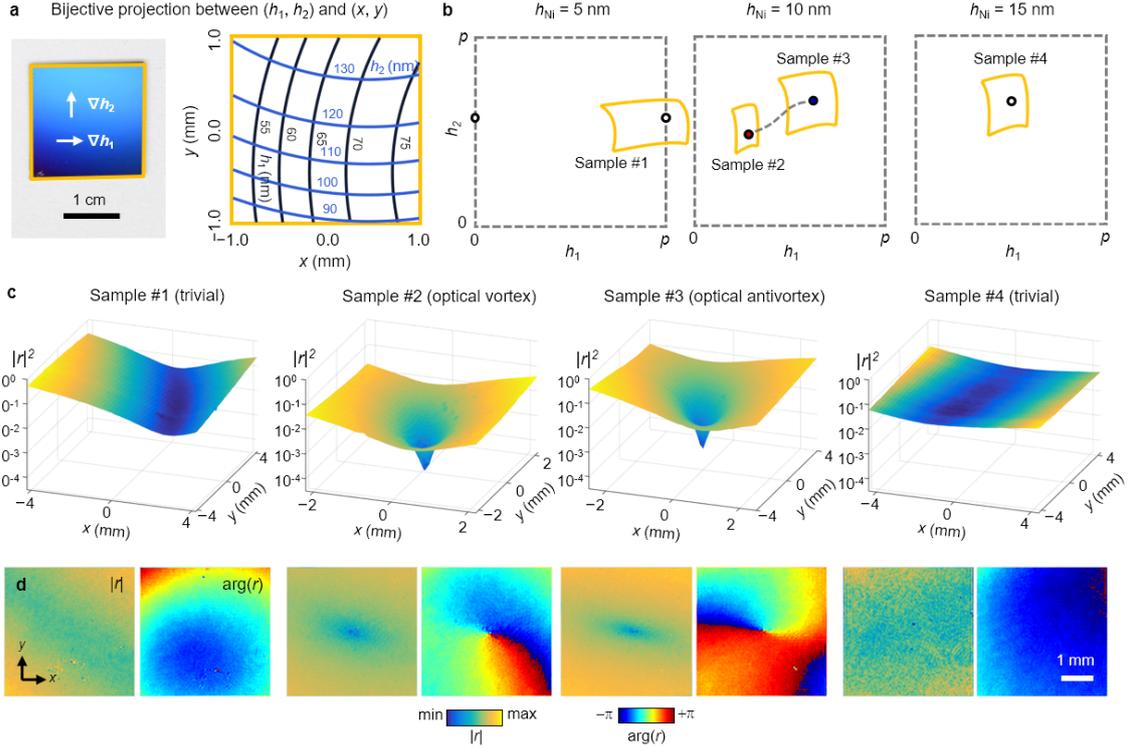

**Figure 2. Measured trivial and non-trivial topological phases in the GTOC. (a)** Picture of a fabricated GTOC (Sample #2), and the thickness distribution of the top and bottom $SiO_2$ layers measured using site-dependent ellipsometry. The black and blue lines indicate the equal-thickness curves of the top and bottom $SiO_2$ layers. **(b)** Bijection of the areas of the employed GTOCs onto the generalized parameter space ($h_1$, $h_2$). Depending on the thickness of the Ni layer, each sample contains a weak or singular reflection minimum within its area (the solid yellow curved polygon) in the generalized parameter space. We investigated four different GTOCs (Sample #1, #2, #3, and #4). The white circle indicates the weak reflection minimum of the trivial topological texture, and the red and blue circles indicate the singular reflection minima of the optical vortex and antivortex, respectively. **(c)** Measured reflectance distribution of the four GTOCs under normal incidence. The wavelength of light is 633 nm. **(d)** Measured amplitude ($|r|$), and phase ($\arg(r)$) distribution of the complex reflection coefficient for the four GTOCs.

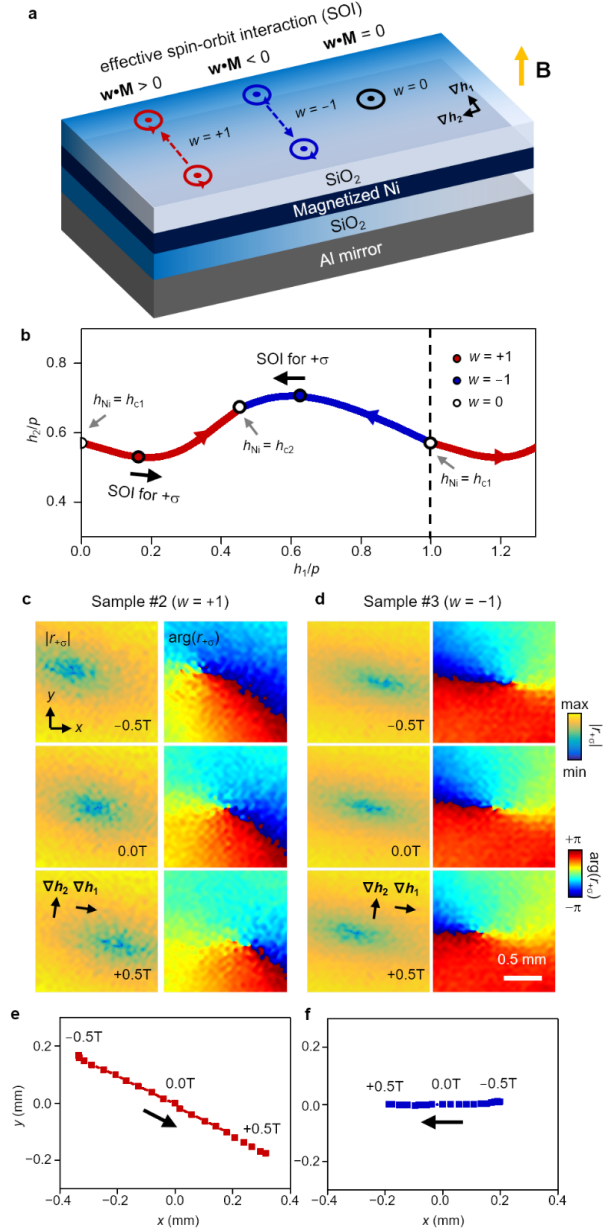

**Figure 3. Effective spin-orbit interaction of the optical vortex and antivortex controllable by the external magnetic field.** (**a**) Schematic illustration of the effective spin-orbit interaction in the GTOC. The spin-orbit interaction (**w·M**) characterizes the dynamics of the optical vortex, antivortex, and weak reflection minimum depending on the topological charge (**w** = $w\hat{z}$, $w$ = +1, −1, and 0), and the magnetization (**M**) of the Ni layer induced by the external magnetic field. (**b**) Calculated trajectory of the singular reflection minima in the generalized parameter space. As the optical thickness of the Ni layer increases, the optical vortex (red circle, $w$ = +1) and antivortex (blue circle, $w$ = −1) emerge at the position of the weak reflection minimum (white circle, $w$ = 0, $h_{Ni} = h_{c1}$), move along the different trajectories, and disappear at the position of the other weak reflection minimum ($h_{Ni} = h_{c2}$), and vice versa. (**c, d**) External magnetic field-dependent behaviour of the complex reflection coefficient ($r_{+\sigma}$) of the optical vortex and antivortex that Sample #2 and #3 (the GTOCs of $h_{Ni}$ = 10 nm in Figure 2) support, respectively. (**e, f**) Trajectories of the optical vortex and antivortex depending on the applied magnetic field.

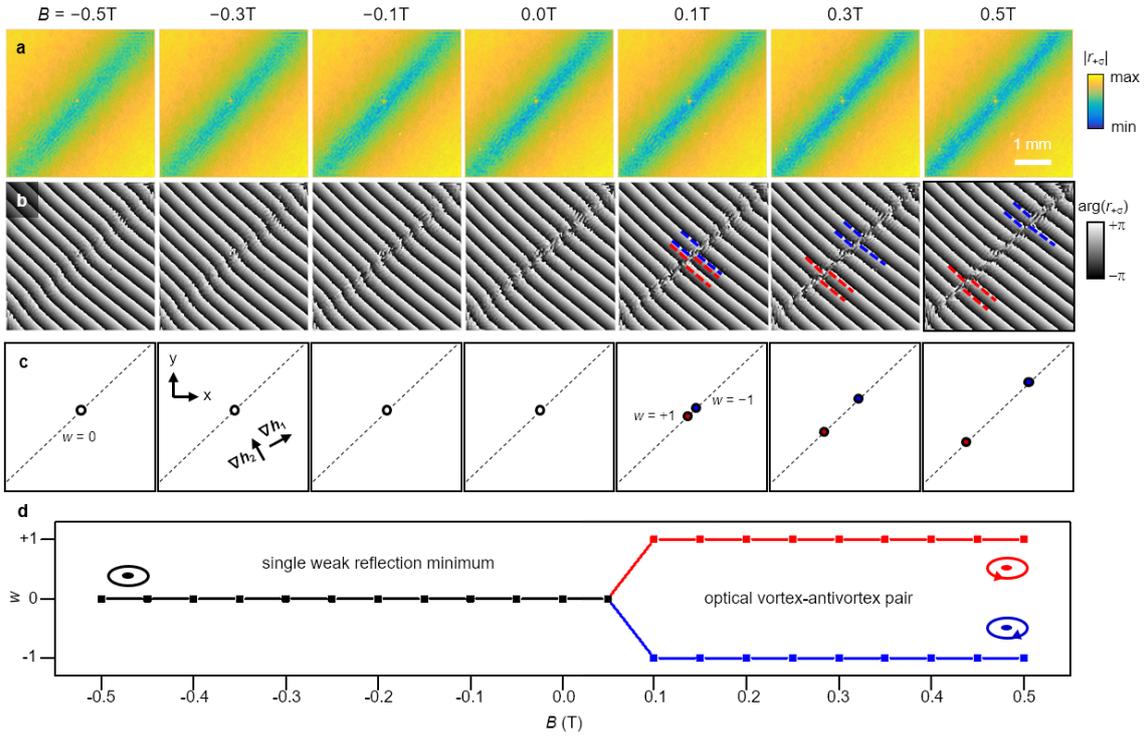

**Figure 4. Magnetically induced generation and annihilation of optical vortex-antivortex pair. (a, b)** Measured reflection **(a)** amplitude, and **(b)** phase distributions of the GTOC (Sample #5, $h_{Ni}$ = ~13 nm) depending on the applied magnetic field. The phase distribution with an added continuously varying global phase clearly identifies the fork patterns of the optical vortex (dashed red guideline, $w$ = +1), and antivortex (dashed blue guideline, $w$ = −1). **(c)** Trajectory of the topological texture depending on the magnetic field. The white circle indicates the position of the weak reflection minimum. The dashed guideline indicates the opposite-direction movements of the optical vortex (red circle) and antivortex (blue circle). **(d)** Topological phase transition controlled by the external magnetic field. The sudden transition between the trivial (single weak reflection minimum) and non-trivial (optical vortex-antivortex pair) topological phases appears at the critical magnetic field of approximately 0.05 T.